# Survey on Incremental Approaches for Network Anomaly Detection

Monowar Hussain Bhuyan[1], D K Bhattacharyya[1] and J K Kalita[2]

[1]Dept. of Computer Science and Engineering[†], Tezpur University, Napaam, Tezpur-784028, Assam, India
[2]Dept. of Computer Science, University of Colorado at Colorado Springs, CO 80933-7150, USA
{[1]mhb, [1]dkb}@tezu.ernet.in, [2]jkalita@uccs.edu

*Abstract*: As the communication industry has connected distant corners of the globe using advances in network technology, intruders or attackers have also increased attacks on networking infrastructure commensurately. System administrators can attempt to prevent such attacks using intrusion detection tools and systems. There are many commercially available signature-based Intrusion Detection Systems (IDSs). However, most IDSs lack the capability to detect novel or previously unknown attacks. A special type of IDSs, called Anomaly Detection Systems, develop models based on normal system or network behavior, with the goal of detecting both known and unknown attacks. Anomaly detection systems face many problems including high rate of false alarm, ability to work in online mode, and scalability. This paper presents a selective survey of incremental approaches for detecting anomaly in normal system or network traffic. The technological trends, open problems, and challenges over anomaly detection using incremental approach are also discussed.

## 1. Introduction

The Internet connects users and providers of information and media services from distant corners of the world. Due to widespread availability of advanced networking technologies, the threat from spammers, intruders or attackers, and criminal enterprises is also increasing. Intrusion Detection Systems (IDSs) and firewall technologies can prevent some of these threats. One study [1] estimates the number of intrusion attempts over the entire Internet to be in the order of 25 billion per day and increasing. McHugh [2] claims that attacks are becoming more sophisticated while they get more automated, and thus the skills needed to launch them are being reduced.

There are two types of IDSs: signature-based and anomaly-based. Signature-based IDSs exploit signatures of known attacks. Such systems require frequent updates of signatures for known attacks and cannot detect unknown attacks or anomalies for which signatures are not stored in the database. In contrast, anomaly based IDSs are extremely effective in finding and preventing known as well as unknown or zero day attacks [3][54]. However, an anomaly detection system has many shortcomings such as high rate of false alarm, and the failure to scale up to gigabit speeds.

The main task of an incremental method is to construct or model the pattern classes and use them for correct classification of the input pattern as either belonging to a certain previously observed class or not belonging to any of referred classes. However, in network anomaly detection, the incremental approach updates normal profiles dynamically based on changes in network traffic without fresh training using all data for attack detection. Based on existing profiles, it can take the decision to raise an attack if abrupt changes happen in normal system or network traffic. Thus, it is useful to detect anomalies from normal system or network traffic from existing signatures or profiles, using an incremental approach. That is, an incremental approach updates profiles or signatures dynamically incorporating new profiles as it encounters them. It does not need to load the whole database each time into the memory and learn fresh from beginning. In the last two decades, a good number of anomaly based intrusion detection approaches [3][4][5][55] have been developed, but a lot of them are general in nature, and thus quite simple. Due to the lack of papers that discuss various facets of incremental anomaly detection, we present a survey in a structured manner along with current research issues and challenges in this field.

This paper reports a selective and comprehensive survey of incremental approaches for network anomaly detection. In Section 2, we introduce the basic problem formulation while Section 3 provides the idea of anomaly detection. A good number of incremental techniques and validity measures for anomaly detection are discussed in Section 4. Section 5 is dedicated to evaluation datasets, whereas Section 6 presents performance evaluation mechanisms. Section 7 discusses research issues and challenges and finally reports concluding remarks in Section 8.

## 2. Problem Definition

Intrusions or attacks can be detected based on the collection of information from a network or a host (normally, in and out traffic). This problem has been formulated in [6] from the pattern recognition view point. Incremental approaches are used to make the system faster in terms of training as well as testing of the instances. There are enormous problems in developing incremental techniques for network anomaly detection due to the dynamic updation of normal as well as attack profiles. We observe that the main importance of incremental network anomaly detection lies in dynamic profile updation for both normal and attack, reduced memory utilization, faster and higher detection rate, and improved real time performance. To address all these issues, we present a comprehensive survey of incremental approaches for network anomaly detection.

## 3. Preface to Anomaly Detection

Anomaly detection refers to the important problem of finding non-conforming patterns or behaviors in live traffic data. These non-conforming patterns are often known as anomalies, outliers, discordant observations, exceptions, aberrations, surprises, peculiarities or contaminants in different application domains. In practice, it is very difficult to precisely detect anomalies in network traffic or normal data. So, anomaly is an interesting pattern due to the effect of traffic or normal data while noise consists of non-interesting patterns that hinder traffic data analysis.

[†]Department of Computer Science & Engineering, Tezpur University is funded by DRS Phase-I under SAP of UGC, Govt. of India.

The central premise of anomaly detection is that intrusive activity is a subset of anomalous activity [3][7]. When there is an intruder who has no idea of the legitimate user's activity patterns, the probability that the intruder's activity is detected as anomalous should be high. Kumar and Stafford [4] suggest four possibilities in such a situation, each with a non-zero probability.

- *Intrusive but not anomalous*: An IDS may fail to detect this type of activity since the activity is not anomalous. But, if the IDS detects such an activity, it may report it as a false negative because it falsely reports the absence of an intrusion when there is one.
- *Not intrusive but anomalous*: If the activity is not intrusive, but it is anomalous, an IDS may report it as intrusive. These are called false positives because an intrusion detection system falsely reports intrusions.
- *Not intrusive and not anomalous*: These are true negatives; the activity is not intrusive and should not be reported as intrusive.
- *Intrusive and anomalous*: These are true positives; the activity is intrusive and much be reported as such.

## 4. Existing Incremental Approaches for Network Anomaly Detection

Recently, several papers have focused on modeling and analyzing anomaly based IDSs. Only a few of these present incremental approaches. Based on a comprehensive survey of published incremental anomaly detection approaches, we conclude that most approaches have high rate of false alarm, are non-scalable, and are not fit for deployment in high-speed networks. Anomaly detection techniques can be of three types: *supervised*, *semi-supervised* and *unsupervised*. Apart from the architecture of an incremental ANIDS, some published papers in each of these three categories are discussed in the following in brief.

### 4.1 Architecture of an Incremental ANIDS
In this section, we present a generic architecture of an incremental anomaly based network intrusion detection system (ANIDS) in Figure 1.

#### 4.1.1 Network Traffic Capture
It is useful to analyze network traffic to detect attacks or anomalies. In a capture module, real network traffic is captured by using Libpcap [10] library, an open source C library offering an interface for capturing link-layer frames over a wide range of system architectures. It provides a high-level common Application Programming Interface to the different packet capture frameworks of various operating systems. The offered abstraction layer allows programmers to rapidly develop highly portable applications.

Libpcap defines a common standard format for files in which captured frames are stored, also known as the tcpdump format, currently a de facto standard used widely in public network traffic archives. Modern kernel-level capture frameworks on UNIX operating systems are mostly based on the BSD (or Berkeley) Packet Filter (BPF) [11][12]. The BPF is a software device that taps network interfaces, copying packets into kernel buffers and filtering out unwanted packets directly in interrupt context. Definitions of packets to be filtered can be written in a simple human readable format using Boolean operators and can be compiled into a pseudo-code to be passed to the BPF device driver by a system call. The pseudo-code is interpreted by the BPF Pseudo-Machine, a lightweight, high-performance, state machine specifically designed for packet filtering. Libpcap also allows programmers to write applications that transparently support a rich set of constructs to build detailed filtering expressions for most network protocols. A few Libpcap calls these Boolean expressions, which can read directly from user's command line, compile into pseudo-code and passed to the Berkeley Packet Filter. Libpcap and the BPF interact to allow network packet data to traverse several layers to finally be processed and transformed into in capture files (i.e., tcpdump format) or in samples for statistical analysis.

#### 4.1.2 Preprocessor
In order to evaluate an IDS, an unbiased intrusion dataset in a standard format is required. Generally, the live captured packet contains a lot of raw data; some of them may not be relevant in the context of an IDS. Therefore, filtrations of irrelevant parameters during captures as well as extraction of relevant features from the filtered data are important preprocessing functions of an IDS. Apart from these, data type conversion, normalization and discretization are also useful functions of this module depending on the anomaly detection mechanism used in the IDS.

**a) Feature Extraction:** Feature extraction from raw data is an important step for anomaly based network intrusion detection. The evaluation of any intrusion detection algorithm on real time network data is difficult, mainly due to the high cost of obtaining proper labeling of network connections. The extracted features are of four types [13][14] and are described below in brief.
- *Basic features*: These can be derived from packet headers without inspecting the payload. For example, the following features are basic features: protocol type, service, flag, source bytes, destination bytes;
- *Content based features*: Domain knowledge is used to assess the payload of the original TCP (Transmission Control Protocol) packets. This type includes features such as the number of failed login attempts;
- *Time-based features*: These features are designed to capture properties that mature over a 2-second temporal window. One example of such a feature is the number of connections to the same host over the 2-second interval;
- *Connection-based features*: These features are computed over a historical window estimated over the number of connections, in this case 100, instead of time.

These features are designed to assess attacks, which span intervals longer than 2 seconds. It is well known that features constructed from the data content of the connections are more important when detecting R2L (Remote to Local) and U2R (User to Root) attack types in KDD99 intrusion dataset [14], while time-based and connection based features are more important for detection of DoS (Denial of Service) and probing attack types [15].

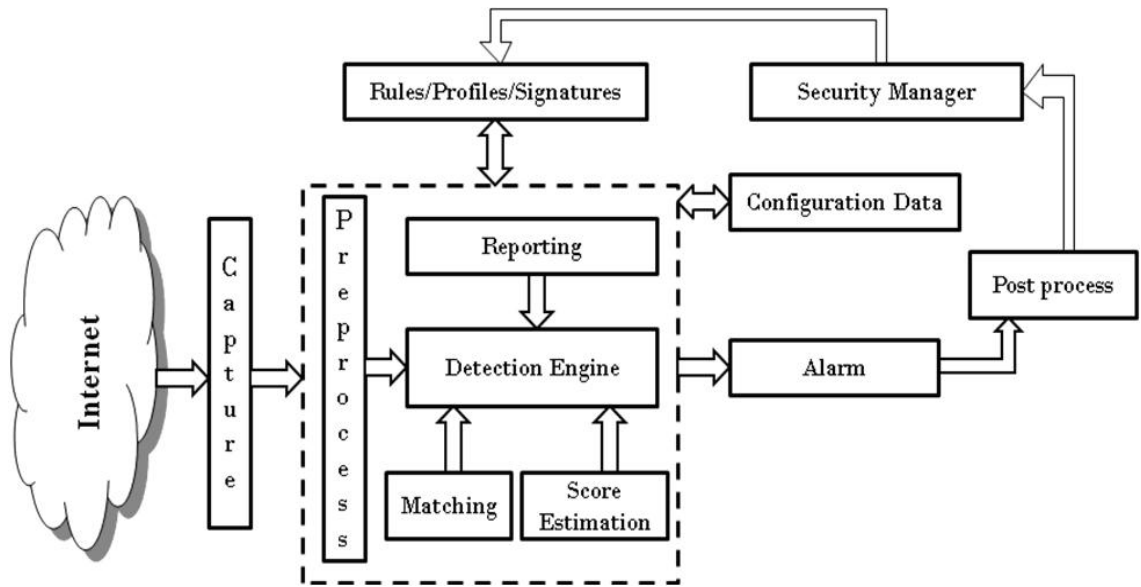

**Figure 1.** A generic architecture of an Incremental ANIDS

**b) Data Type Conversion:** Both features and raw data may include numeric as well as categorical data (e.g., tcp, icmp (Internet Control Message Protocol), telnet, etc). Therefore, to apply a clustering anomaly detection technique based on a proximity measure for numeric or categorical data, it may be necessary to convert the data.

**c) Normalization:** In an intrusion dataset, all parameters or field values may not be equally weighted. In such cases, normalization is considered useful before applying an anomaly detection mechanism.

**d) Discretization:** The network intrusion data contains several continuous valued attributes such as number of packets, number of bytes, duration of each connection, etc. These attributes may need to be transformed into binary features before applying any standard association mining algorithms. The transformation can be performed using a variety of supervised and unsupervised discretization techniques. For example, using the output scores of the anomaly detector as its ground truth, MINDS (Minnesota INtrusion Detection System) [16] employs a supervised binning strategy to discretize attributes. Initially, all distinct values of a continuous attribute are put into one bin. The worst bin in terms of purity is selected for partitioning until the desired number of bins is reached. The discretization of numeric attributes contributes to the comprehension of the final results.

*4.1.3 Anomaly Detection Engine*

Anomaly detection is usually flexible and adaptive enough to detect novel attacks. There are still many problems in anomaly detection approaches. Some issues in anomaly detection are discussed below.

**a) Types of Anomalies:** The purpose of anomaly detection is to find anomalies in various application domains. An anomalous MRI image may indicate presence of malignant tumors [17]. Anomalies in credit card transaction data may indicate credit card or identity theft [18]. Anomalous readings from a spacecraft sensor can signify a fault in some component of the spacecraft [19]. Detection of anomalies or outliers were discussed as early as in the 9th century [20]. Anomalies are classified into three categories based on [5].

- *Point anomalies*: An individual data instance may be anomalous with respect to the rest of the data.
- *Contextual anomalies*: A data instance may be anomalous in a specific context with respect to a condition [21]. Two sets of attributes are normally used to detect this type of anomalies: (a) contextual attributes and (b) behavioral attributes. However, Song et al. [21] use the terms environmental and indicator attributes respectively against a context. Contextual attributes determine the context or neighborhood of an instance. Behavioral attributes are responsible for the non-contextual characteristics of an instance. In known point anomaly detection techniques, identifying a context and computing anomaly score for each test data instance are generic steps.

To detect these anomalies, Song et al. [21] propose a model which is based on contextual and behavioral attributes. The authors assume that the attributes are pre-partitioned. Each test data instance $d$ is represented as $(x, y)$. $U$ is the contextual data instance, which is partitioned by using Gaussian Mixture Model (GMM). $V$ is the behavioral data instance, which is also partitioned by using another Gaussian Mixture Model (GMM). $P(V_j | U_i)$ is a mapping function indicates that the probability of the indicator part of data point y to be generated from $V_j$, whereas the environmental part $x$ is generated by using $U_i$. Thus the anomaly score, $AS_d$ for a given test data instance $d = (x, y)$ is given by-

$$AS_d \sum_{i=1}^{n_U} p(x \in U_i) \sum_{j=1}^{n_V} p(y \in V_j) p(V_j | U_i)$$

where $n_U$ is the number of mixture components in $U$ and $n_V$ is the number of mixture components in $V$. $p(x \in U_i)$ is the probability of a sample point $x$ to be generated from the mixture component $U_i$ while $p(y \in V_j)$ is the probability of a sample point $y$ to be generated from the mixture component $V_j$.

- *Collective anomalies*: Sometimes, it is not a single data point that is anomalous, but it is a collection of related data instances that is anomalous with respect to the entire dataset. It is worth mentioning that unlike point anomalies, detection of collective anomalies requires identifying related anomalous points in a dataset. To detect contextual anomalies, one needs to identify the appropriate contextual or behavioral attributes in the data.

**b) Data Labels:** Labeling a data instance as normal or anomalous based on its behavior is often prohibitively expensive. A human expert often does labeling manually and hence it requires substantial effort to obtain the labeled training dataset. Typically, getting a labeled set of anomalous data instances covering all possible types of anomalous behavior is more difficult than getting labeled data for normal behavior.

**c) Reporting of anomalies:** An important aspect for any anomaly detection technique is the way it reports anomalies. Typically, the outputs produced by anomaly detection techniques are one of the following two types.

- *Scores*: An anomaly score is assigned to each instance in the test data depending on the degree to which that instance is considered an anomaly. Usually, the output of such a technique is a ranked list of anomalies. An analyst may choose to either analyze the top few anomalies or use a threshold to select anomalies for further investigation [22].
- *Labels*: A label (normal or anomalous) is assigned to each test instance. Scoring based anomaly detection techniques allow the analyst to use a domain specific threshold to select the most relevant anomalies. Generally, an IDS uses clustering techniques for grouping activities. It merely classifies the data, without any interpretation of the data. A classical labelling strategy is to measure the cardinality of the clusters, and label some percentage of the smallest clusters as malicious. This approach does, however, have some limitations, and does not detect massive attacks [23] properly, e.g. Denial-of-Service attacks. Dunn's index [24] or C-index [25] are well suited for clustering quality evaluation indices. A hallmark of good clustering quality is compact clusters distant from each other. Dunn's index (D) is defined as the ratio between the minimal intra-cluster distances $d_{min}$ to maximal inter-cluster distance $d_{max}$, i.e., $d_{min} / d_{max}$. $D$ is limited to the interval [0, 1] and should be maximized for good clustering. On the other hand, C-index (C) is defined in terms $S$, i.e., the sum of distances over all pairs of objects in the same cluster. Let n be the number of such object pairs, and $S_{min}$ be the sum of the $n$ smallest distances if all pairs of objects are considered. Likewise, let $S_{max}$ be the sum of the $n$ largest distances out of all pairs. Then $C$ is computed as $(S - S_{min}) / (S - S_{max})$. $C$ is restricted to the interval [0, 1] as well and should be minimized for good clustering. $D$ measures only two distances whereas $C$ requires clusters of equal cardinality to produce proper quality evaluations.

**d) Matching:** Matching is the most important functional module in the incremental anomaly based intrusion detection architecture. Generally, it can be done either, by using a distance based approach, or, by a density based approach, or by a combination of the both. In a distance based approach, the proximity measure plays a vital role as in classification, clustering, and information retrieval. From scientific and mathematical points of view, distance is defined as a quantitative degree of how far apart two objects are [26]. A synonym for distance is dissimilarity. Distance measures satisfying metric properties are simply called metric while other non-metric distance measures are occasionally called divergence. A synonym for similarity is proximity. Similarity measures are often called similarity coefficients. The choice of distance or similarity measures depends on the measurement type for the object's attributes or the representation of objects. However, proximity-based approaches typically take $O(n^2)$ matching time. On the other hand, a density based approach attempts to match by using the nearest neighbour concept. It is closely related to the proximity based approach since density is defined in terms of proximity. One common approach is to define density as the reciprocal of the average distance to the k-nearest neighbours [27]. If the distance is small, the density is high, and vice versa. Anomalies in the test data are usually found in the low density areas. If the density is high, the occurrence of anomalies is low and vice versa. However, it takes $O(n^2)$ time for matching. It can be reduce up to $O(n\log n)$ by using appropriate data structures.

### *4.1.4 Rule, Profile or Signature*

In misuse based intrusion detection, rules, profiles or signatures play a very important role. It is generated based on true normal data or on existing or known attack data. However, updation of such a rule or profile or signature base dynamically is very essential to cope with evolving network scenarios. But due to the lack of availability of exhaustive true normal data, generating of an effective rule-base or set of profiles is still a challenging task.

### *4.1.5 Security Manager*

In this phase, the security manger handles the rule, profile or signature base to update dynamically based on the feedback obtained from the arrival of new type of normal data or on the detection of new type of attacks. We introduce post processing of alarms to reduce the number of false positives.

### *4.1.6 Alarm*

An alarm is generated based on the anomaly score or threshold assigned by the security manager.

### **4.2 Supervised Approaches**

In this approach, a predictive model is developed based on a training dataset (i.e., data instances labeled as normal or attack class). Any unseen data instance is compared against the model to determine which class it belongs to. There are two major issues that arise in supervised anomaly detection. First, the anomalous instances are far fewer in number compared to normal instances in the training data. Issues that arise due to imbalanced class distributions have been addressed in the data mining and machine learning literature [8]. Second, obtaining accurate and representative labels, especially for the anomaly class is usually challenging. A number of techniques have been proposed that inject artificial anomalies in a normal dataset to obtain a labeled training dataset [9]. Other than these two issues, the supervised anomaly detection problem is similar to building predictive models. We now discuss some of the most common incremental supervised anomaly detection approaches.

1) **Yu et al. [28]:** The authors propose an incremental learning method by cascading a service classifier (SC) over an incremental tree inducer (ITI) for supervised anomaly detection. A service classifier ensures that the ITI method is trained with instances with only one service value (e.g., ftp, smtp, telnet, etc.). The ITI method is trained with instances incrementally without the service attribute. The cascading method has three phases: (i) training, (ii) testing and (iii) incremental learning. During training phase, the service classifier method is first applied to partition the training dataset into $m$ disjoint clusters according to different services. Then, the ITI method is trained with the instances in each cluster. The service classifier method ensures that each training instance is associated with only one cluster. In the testing phase, it finds the clusters to which the test instances belong. Then, the ITI method is tested with the instances. In the incremental learning phase, the service classifier and ITI cascading method are not re-trained; the authors use incremental learning to train the existing ITI binary tree. Nearest Neighbor combination rules embedded within K-Means+ITI and SOM (Self-Organizing Map)+ITI cascading methods are used in experiments. The authors also compare the performance of SC+ITI with K-Means, SOM, ITI, K-Means+ITI and SOM+ITI methods in terms of detection rate and false positive rate (FPR) on the KDD'99 dataset. Results show that the ITI method has better performance than K-Means, SOM, K-Means+ITI and SOM+ITI methods in terms of overall detection rate. However, the SC+ITI cascading method outperforms the ITI method in terms of detection rate and FPR, and obtains better detection rate compared to other methods. Like the ITI method, SC+ITI also provide additional options for handling missing values and incremental learning.

2) **Lascov et al. [29]:** The authors focus on the design and analysis of efficient incremental SVM learning, with the aim of providing a fast, numerically stable and robust implementation. A detailed analysis of convergence and algorithmic complexity of incremental SVM learning is carried out in this paper. Based on this analysis, a new design for storage and for numerical operations is proposed. The design speeds up the training of an incremental SVM by a factor of 5 to 20. The performance of the new algorithm is demonstrated in two scenarios: learning with limited resources and active learning. The authors discuss various applications of the algorithm, such as in drug discovery, online monitoring of industrial devices and surveillance of network traffic.

3) **Ren et al. [30]:** The authors propose a new anomaly detection algorithm that can update the normal profile of system usage dynamically. The features used to model a system's usage pattern are derived from program behavior. A new program behavior is inserted into old profiles by density-based incremental clustering when system usage pattern changes. It is much more efficient compared to traditional updating by re-clustering. The authors test their model using the 1998 DARPA BSM audit data, and report that the normal profiles generated by their algorithm are less sensitive to noise data objects than profiles generated by the ADWICE algorithm. The method improves the quality of clusters and lowers the false alarm rate.

4) **Khreich et al. [52]:** The authors propose a system based on the receiver operating characteristic (ROC) to efficiently adapt ensembles of HMMs (EoHMMs) in response to new data, according to a learn-and-combine approach. When a new block of training data becomes available, a pool of base HMMs is generated from the data using a different number of HMM states and random initializations. The responses from the newly-trained HMMs are combined to those of the previously-trained HMMs in ROC space using a novel incremental Boolean combination (incrBC) technique. Finally, specialized algorithms for model management allow selecting a diversified EoHMM from the pool, and adapting Boolean fusion functions and thresholds for improved performance, while it prunes redundant base HMMs. The proposed system is capable of changing the desired operating point during operations, and those points can be adjusted to changes in prior probabilities and costs of errors. The authors' simulations on synthetic and real-world host-based intrusion detection data indicate that the proposed system can achieve a significantly higher level of performance than when parameters of a single best HMM are estimated, at each learning stage, using reference batch and incremental learning techniques. It also outperforms the learn-and-combine approaches using static fusion functions (e.g., majority voting). Over time, the proposed ensemble selection algorithms form compact EoHMMs, while maintaining or improving system accuracy. Pruning allows limiting the pool size from increasing indefinitely, thereby reducing the storage space for accommodating HMM parameters without negatively affecting the overall EoHMM performance. Although applied for HMM-based anomaly detection systems, their approach is general and can be employed for a wide range of classifiers and detection applications.

5) **Lu et al. [53]:** The authors use Commute Time Distance (CTD) as a metric for anomaly detection. It is a random walk based metric on graphs. They show that CTD can be used to simultaneously identify both global and local anomalies. They propose an accurate and efficient approximation for computing CTD in an incremental fashion to facilitate real-time applications. They develop an online anomaly detection algorithm, where the CTD of each new arriving data point to any point in the current graph can be estimated in constant time ensuring a real-time response.

6) **Yi et al. [56]:** The authors develop RS-ISVM, an improved incremental SVM algorithm to deal with network intrusion detection. To reduce the noise generated by feature differences, the authors propose a modified kernel function U-RBF, with the mean and mean square difference values of feature attributes embedded in the RBF kernel function. Given the oscillation problem that usually occurs in traditional incremental SVM's follow-up learning process, they present a reserved set strategy which can keep those samples that are more likely to be the support vectors in the computation process. Moreover, in order to shorten the training time, a concentric circle method is suggested to be used in selecting samples to form the reserved set. The authors claim that RS-ISVM can ease the oscillation phenomenon in the learning process and achieve good performance while its reliability remains relatively high.

### 4.3 Semi-supervised Approaches

In semi-supervised approach, the training data instances belong to the normal class only. Data instances are not labeled for the attack class. There are many approaches used to build the model for the class corresponding to normal behavior. This model is used to identify anomalies in the test data. Some of the detection methods are discussed in the following.

1) **Burbeck et al. [31]:** ADWICE (Anomaly Detection With fast Incremental Clustering) uses the first phase of the BIRCH clustering framework [32] to implement fast, scalable and adaptive anomaly detection. It extends the original clustering algorithm and applies the resulting detection mechanism for analysis of data from IP networks. The performance is demonstrated on the KDD99 intrusion dataset as well as on data from a test network at a telecom company. Their experiments show good detection quality (95%) and acceptable false positives rate (2.8 %) considering the online, real-time characteristics of the algorithm. The number of alarms is further reduced by application of the aggregation techniques implemented in the Safeguard architecture[1].

2) **Rasoulifard et al. [33]:** It is important to increase the detection rate for known intrusions and also to detect unknown intrusions at the same time. It is also important to incrementally learn new unknown intrusions. Most current intrusion detection systems employ either misuse detection or anomaly detection. In order to employ these techniques effectively, the authors propose an incremental hybrid intrusion detection system. This framework combines incremental misuse detection and incremental anomaly detection. The framework can learn new classes of intrusion that do not exist in data used for training. The framework has low computational complexity, and so it is suitable for real-time or on-line learning. The authors use the KDDcup99 intrusion dataset to establish this method.

3) **Kalle et al. [34]:** Anomaly detection is very expensive in real-time. First, to deal with massive data volumes, one needs to have efficient data structures and indexing mechanisms. Second, the dynamic nature of today's information networks makes the characterization of normal requests and services difficult. What is considered normal during some time interval may be classified as abnormal in a new context, and vice versa. These factors make many proposed data mining techniques less suitable for real-time intrusion detection. The authors look at the shortcomings of ADWICE and propose a new grid index that improves detection performance while preserving efficiency in search. Moreover, they propose two mechanisms for adaptive evolution of the normality model: incremental extension with new elements of normal behavior, and a new feature that enables forgetting of outdated elements of normal behavior. It evaluates the technique for network-based intrusion detection using the KDD99 intrusion dataset as well as on data from a telecom IP test network. The experiments yield good detection quality and act as proof-of-concept for adaptation of normality.

### 4.4 Unsupervised Approaches

Unsupervised detection approaches do not require training data, and thus are most widely applicable. These techniques make the implicit assumption that normal instances are far more frequent than anomalies in the test data. If this assumption is not true, such techniques suffer from high false alarm. Most existing unsupervised anomaly detection approaches are clustering based. Clustering is a technique to group similar objects. It deals with finding structure in a collection of unlabeled data. Representing the data by fewer clusters necessarily leads to the loss of certain finer details, but achieves simplification. In anomaly detection, clustering plays a vital role in analyzing the data by identifying various groups as either belonging to normal or to anomalous categories. There are many different clustering based anomaly detection approaches in the literature. Most commonly used clustering techniques are: partitioning-based (e.g., Zhong et al. [35]), hierarchical (e.g., Hsu et al. [36], Burbeck et al. [31], Kalle et al. [34]), density-based (e.g., Ren et al. [30]), and grid-based techniques. These techniques are discussed in the following in brief.

1) **Hsu et al. [36]:** Adaptive resonance theory network (ART) is a popular unsupervised neural network approach. Type I adaptive resonance theory network (ART1) deals with binary numerical data, whereas type II adaptive resonance theory network (ART2) deals with general numerical data. Several information systems collect mixed type attributes, which include numeric attributes and categorical attributes. However, both ART1 and ART2 do not deal with mixed type data. If the categorical attributes are transferred to binary data format, the binary data do not reflect reality with the same fidelity. It ultimately influences clustering quality. The authors present a modified adaptive resonance theory network (M-ART) and a conceptual hierarchy tree to solve the problem of clustering with mixed data. They show that the M-ART algorithm can process mixed type data well and has a positive effect on clustering.

---

[1]Safeguard: The safeguard project, (online) http://www.safeguardproject.info/

2) **Zhong et al. [35]:** This paper presents an incremental clustering algorithm for intrusion detection using clonal selection based on a partitioning approach. It partitions the dataset into initial clusters by comparing the distance from data to cluster centroid with the size of cluster radius, and analyzes the clustering data with mixed attributes by using an improved definition of distance measure. The objective function optimizes clustering results by applying a clonal selection algorithm [37], and then labels clusters as normal or anomalous as appropriate. The authors try to find better cluster centroids to improve the partitioning quality which is evaluated by the objective function. If the value of objective function is small, the sum of the distances from data to the cluster centers is also small and the objects in the same cluster are more close to each other. The method attempts to optimize cluster results from one iteration to the next using the clonal selection algorithm [37]. The authors establish this incremental technique in terms of high detection rate and low false positive rate.

### 4.5 Discussion

Based on this short and selective survey of incremental techniques for anomaly detection, we make the following observations.

- Most incremental anomaly detection techniques have been benchmarked by using KDDcup99 intrusion dataset; hence, they cannot claim to be truly up-to-date. It is an offline dataset. The KDDcup99 intrusion datasets have been already established to be biased considering the presence of normal and attack data in unrealistic proportions [22]. So, the approaches need to be tested over real time intrusion dataset.
- Performances of these techniques are not great in terms of detection rates. Of the existing techniques, ADWICE grid [34] has the highest detection rate, and Hybrid IDS [33] has the lowest detection rate as shown in Table 2. However, the techniques were tested with offline dataset only.
- Most techniques [36][35][31][34][30] use clustering for anomaly detection as shown in Table 1, although other techniques are also used. Clustering is popular because it is easy to create profiles based on the clusters. Most techniques use Euclidean distance as the proximity measure. The detection rates fairly high.
- We observe that most techniques are non-real time and use distance or cluster based features for profile generation, though there are a few real time systems [29][31][34] as claimed by the authors.
- We have identified pros and cons of each approach with performance in terms of detection rate and false positive rate these are seen in Table 2.

### 4.6 Validity Measures

The performance of any ANIDS is highly dependent upon (i) its individual configuration, (ii) the network it is monitoring and (iii) its position within that network [38]. Simply benchmarking an ANIDS once in a certain environment does not provide a definitive method for assessing it in a given situation. A number of metrics [38] are used to assess the suitability of a given ANIDS for a particular situation or environment. Some of these are discussed next.

- *Ability to identify attacks*: The main performance requirement of an ANIDS is to detect intrusions. In particular, many vendors and researchers appear to consider any attempt to place malicious traffic on the network as an intrusion. In reality a more useful system will log malicious traffic and only inform the operator if the traffic poses a serious threat to the security of the target host.
- *Known vulnerabilities and attacks*: All ANIDSs should be capable of detecting known vulnerabilities. However research indicates that many commercial IDSs fail to detect recently discovered attacks. On the other hand if a vulnerability or attack becomes known, all systems should be patched; otherwise, it should be applied so that the need for an ANIDS to detect these events is obviated.
- *Unknown attacks*: Detecting unknown attacks must be the most important feature of any ANIDS. Only the ability of an ANIDS to detect attacks that are not yet known justifies expenses of its implementation and deployment. New vulnerabilities are being discovered every day and being able to detect known attacks is no longer enough.
- *Stability, Reliability and Security*: An ANIDS should be able to continue operating consistently in all circumstances. The application and the operating system should ideally be capable of running for months, even years without segmentation faults or memory leakage. An important requirement imposed on an ANIDS is the ability to consistently report identical events in the same manner. The system should also be able to withstand attempts to compromise it. The ability of an attacker to identify an ANIDS on a network can be extremely valuable to the attacker. The attacker may then attempt to disable the ANIDS using DoS or DDoS techniques. The ANIDS system should be able to withstand all of these types of attack.
- *Identifying target and source*: An alert raised after detecting an anomaly should also identify the source of the threat and the exact identity of the target system. Additional information from whois or DNS lookup on an IP address should also be obtained, if necessary.
- *Outcome of attack*: Another useful feature of an ANIDS should be to determine the outcome of an attack (success or failure). In most cases, an alert simply indicates that an attempt to intrude has been made. It is then the responsibility of the analyst to search for correlated activities to determine the outcome of the attack. If an ANIDS were to present the analyst with a list of other alerts generated by the target host, and a summary of other (non-alert) traffic, the evaluation of the outcome can be greatly accelerated.

- *Legality of data collected*: The legality of the data collected by an ANIDS is of extreme importance if any legal activity may be pursued against the attacker. A disturbingly large number of systems do not collect the actual network packets; instead they simply record their own interpretation of events. A more robust system must capture and store network traffic in addition to simply raising the alert.
- *Signature updates*: An ANIDS should have the ability to detect new types of intrusions and effectively update signatures dynamically.

**Table 1.** General comparison of various incremental approaches for network anomaly detection

| Detection approaches | Method | Real (R)/ Non-real (N) Time | Basic Idea | Proximity Measure | Clustering / SVM / DTree/ Hybrid | Dataset used | Data Type |
|---|---|---|---|---|---|---|---|
| Supervised | SC+ITI based [28] | N | cluster based service value | Nearest Neighbor | Decision Tree | KDDCup99 | Numeric |
| | SVM [29] | R | objective function | - | SVM | KDDCup99 | Numeric |
| | Density-based [30] | N | Distance, No. of points within r-radius | Nearest Neighbor | Clustering | DARPA98 BSM | Numeric |
| | EoHMMs [52] | N | HMMs | Log likely-hood function | Hybrid | UNM datasets | Numeric |
| | RS-ISVM [56] | N | RBF based kernel function | - | SVM | KDDcup99 | Numeric |
| | iECT [53] | R | Laplacian matrix | Compute time distance | Hybrid | DBLP, KDDcup99, and NICTA | Numeric |
| Semi-supervised | ADWICE TRD [31] | R/N | CF (cluster feature) tree | Euclidean Distance | Clustering | KDDCup99 | Numeric |
| | Hybrid IDS [33] | N | weighted majority voting | Average Distance | Hybrid | DARPA98 | Numeric |
| | ADWICE Grid [34] | R/N | CF tree based index | Euclidean Distance | Clustering | KDDCup99 | Numeric |
| Unsupervised | M-ART [36] | N | distance-based tree | LCP (least common points) | Clustering | UCI | Mixed |
| | Clonal Selection [35] | N | distance | Manhattan Distance | Clustering | KDDCup99 | Mixed |

## 5. Evaluation Datasets

Researchers use various datasets for testing and evaluating different anomaly detection methods. It is very difficult to evaluate an anomaly detection system based on live network traffic or any raw network traffic. That is why some benchmark datasets are used for evaluating an anomaly detection system. Some of these are discussed in brief.

### 5.1 Lincoln Lab Datasets

In 1998, MIT's Lincoln Lab performed an evaluation of anomaly-based intrusion detection systems [39]. To perform this evaluation, the Lab generated the DARPA training and testing datasets. Both datasets contain attacks and background traffic. In 1999, the KDDCup competition used a subset of the preprocessed DARPA training and test data supplied by Solvo and Lee [40], the principal researchers for the DARPA evaluation. The raw training data was about four gigabytes of compressed binary tcpdump[2] data from seven weeks of network traffic. This
was processed into about five million connection records. The dataset is known as the KDD99 dataset and a summary of it is reported in Table 7 [41][42][43]. There are four main types of attacks: denial of service, remote-to-local, user-to-root, and surveillance or probing. Background traffic is simulated and the attacks are all known. The training set, consisting of seven weeks of labeled data, is available to the developers of intrusion detection systems. The testing set also consists of simulated background traffic and known attacks, including some attacks that are not present in the training set. We discuss the four categories of attack patterns in details next.

### 5.1.1 Denial of service attacks

This class of attacks makes some computing or memory resource too busy or too full to handle legitimate requests. Thus legitimate users' access are denied. There are several ways to launch DoS attacks such as: by abusing the computers' legitimate features, by targeting the implementations bugs, by exploiting the system's misconfigurations. It is classified based on the services that an attacker renders unavailable to legitimate users. Some of the popular DoS attacks with characteristics are shown in Table 3.

### 5.1.2 Probing attacks

Probing is a class of attacks, where an attacker scans a network to gather information or find known vulnerabilities or finds the weaknesses of the machines. An attacker tries to exploit the attacks, when weaknesses of one or more machines have been found. There are different types of probe attacks. Some of them abuse a computer's legitimate features, some of them use social engineering techniques, and some of them use weaknesses

---

[2](online) http://www.tcpdump.org/

of the machine. A set of probe attacks are shown in Table 4.

### 5.1.3 User to root attacks

In user to root exploits, the attacker starts out with access to a normal user account on the system and is able to exploit vulnerability to gain root access to the system. Most common exploits in this class of attacks are regular buffer overflows, which are caused by regular programming mistakes and environment assumptions. Following are the list of user to root attacks (refer to Table 5).

**Table 2.** Pros and Cons of the methods

| Method | Summary | | Performance | |
| --- | --- | --- | --- | --- |
| | Pros | Cons | Detection Rate (%) | False Positive Rate (%) |
| SC+ITI based [28] | • Performs well over force assignment problem and class dominance problem in terms of detection rate and false positive rate. | • Performance comparison with other relevant techniques missing.<br>• Not tested in a real-time environment. | 92.63 | 1.80 |
| SVM [29] | • Tested their approach in two possible scenarios: (*i*) learning with limited resources and (*ii*) active learning. They claim that it performs in constant running time during learning as well as prediction. | • Too many input parameters. | - | - |
| Density-based [30] | • Reduces the cost of re-clustering and profile updating and claims to achieve higher accuracy than ADWICE. | • In parameter level, it does not make any improvement.<br>• Lack of performance results over larger datasets. | - | - |
| EoHMMs [52] | • On synthetic and real-world HIDS datasets, achieves higher detection accuracy when single best HMM are estimated at each learning stage. | • Not explored the dynamic selection of the best classifiers, decision threshold, Boolean fusion function during the operations. | - | - |
| RS-ISVM [56] | • Reduces the training time as well as prediction time.<br>• Performs better than the simple SVM in terms of detection rate and false positive rate. | • Performance is poor in case of U2R and R2L attacks.<br>• It needs to reduce the number of parameters during training phase. | 89.17 | 4.9 |
| iECT [53] | • Capable to update the eigenvectors and eigenvalues of the graph Laplacian matrix incrementally.<br>• Capable to estimate CTD in constant time incrementally using the property of random walk and hitting time. | • iECT is faster than iLED but it can only be used in case where a new test point is added and cannot be used when there are weigh updates in the graph.<br>• They take only a small subset of datasets for testing. | - | - |
| ADWICE TRD [31] | • Performs better on KDDcup99, DARPA98 and safeguard framework datasets. | • ADWICE-TRD explores the BIRCH cluster features for tree formation but it is not feasible in network anomaly detection. | 95 | 2.8 |
| Hybrid IDS [33] | • Capable to detect unknown attacks i.e., the instance of attacks are not used in training. | • They use only DARPA98 dataset. Hence, it needs to be tested over real-life network intrusion datasets. | 92.63 | 1.8 |
| ADWICE Grid [34] | • Performs better than ADWICE-TRD and ADWICE-TRR in terms of detection rate and false alarm rate on real world datasets in safeguard testbed. | • They claim that their approach adaptively updates the normal profiles but it is important that the profiles need to be updated based on the different network scenarios. | 97.2 | 1.8 |
| M-ART [36] | • Can work well where both numeric and categorical data are available.<br>• Performs well in terms of detection rate on synthetic and real world UCI datasets. | • Not tested on high dimensional large datasets.<br>• Does not work well where it contains only in numeric data values. | 95 | 2.8 |
| Clonal Selection [35] | • Capable to work in mixed type datasets based their modified objective function.<br>• Performs well in terms of detection rate and false positive rate. | • May not work well in real-time scenarios. | 98.3 | 1.8 |

### 5.1.4 Remote to local attacks

A remote to local (R2L) attack makes attack by sending packets to a machine over a network, and then exploits the machines vulnerability to illegally gain local access as a user. There are different types of R2U attacks. The most common attack in this class uses social engineering. Some R2L attacks are reported in Table 6.

### 5.2 LBNL Datasets

This dataset can be obtained from Lawrence Berkeley National Laboratory (LBNL) in the USA. Traffic in this dataset is comprised of packet-level incoming, outgoing, and internally routed traffic streams at the LBNL edge routers. Traffic was anonymized using the tcpmkpub tool [44]. The main applications observed in internal and external traffic are Web (i.e., HTTP), Email, and Name Services. The dataset identify attack traffic by isolating the corresponding scans in aggregate traffic traces. The outgoing TCP scans in the dataset follow LBNL hosts for resetting the TCP connection. Clearly, the attack rate is significantly lower than the background traffic rate (see Table 8 for detailed statistics).

### 5.3 Endpoint Datasets

The traffic rates observed at the end-points are much lower than those at the LBNL routers. The large traffic volumes of home computers are also evident from their high mean numbers of sessions per second. To generate attack traffic, the analysts infected VMs on the end-points with different malwares: Zotob.G, Forbot-FU, Sdbot-AFR, Dloader-NY, So-Big.E@mm, My-Doom.A@mm, Blaster, Rbot-AQJ, and RBOT.CCC. Details of the malwares can be found at Semantic Security Response website[3]. The attack traffic logged at the end-points is mostly comprised of outgoing port scans. Moreover, the attack traffic rates at the end-points are generally much higher than the background traffic rates of LBNL datasets. For each malware, attack traffic of 15 minutes duration was inserted in the background traffic at each end-point at a random time instance. The background and attack traffic statistics of the end-point datasets [45] are given in Table 9 and Table 10.

**Table 3.** DoS attacks and its characteristics

| Attack type | Service | Mechanism | Effect of the attack |
|---|---|---|---|
| Apache2 | http | Abuse | Crashes httpd |
| Back | http | Abuse/Bug | Slows down server response |
| Land | http | Bug | Freezes the machine |
| Mail bomb | N/A | Abuse | Annoyance |
| SYN flood | TCP | Abuse | Denies service on one or more ports |
| Ping of death | ICMP | Bug | None |
| Process table | TCP | Abuse | Denies new processes |
| Smurf | ICMP | Abuse | Slows down network |
| Syslogd | Syslog | Bug | Kills the Syslogd |
| Teardrop | N/A | Bug | Reboots the machine |
| Udpstorm | Echo/Chargen | Abuse | Slows down the network |

**Table 4.** Probe attacks and its characteristics

| Attack type | Service | Mechanism | Effect of the attack |
|---|---|---|---|
| Ipsweep | ICMP | Abuse of feature | Identifies active machines |
| Mscan | Many | Abuse of feature | Looks for known vulnerabilities |
| Nmap | Many | Abuse of feature | Identifies active ports on a machine |
| Saint | Many | Abuse of feature | Looks for known vulnerabilities |
| Satan | Many | Abuse of feature | Looks for known vulnerabilities |

**Table 5.** User to root attacks and its characteristics

| Attack type | Service | Mechanism | Effect of the attack |
|---|---|---|---|
| Eject | User session | Buffer overflow | Gains root shell |
| Ffbconfig | User session | Buffer overflow | Gains root shell |
| Fdformat | User session | Buffer overflow | Gains root shell |
| Loadmodule | User session | Poor environment sanitation | Gains root shell |
| Perl | User session | Poor environment sanitation | Gains root shell |
| Ps | User session | Poor temp file management | Gains root shell |
| Xterm | User session | Buffer overflow | Gains root shell |

**Table 6.** Remote to local attacks and its characteristics

| Attack type | Service | Mechanism | Effect of the attack |
|---|---|---|---|
| Dictionary | Telnet, rlogin, pop, ftp, imap | Abuse feature | Gains user access |
| Ftp-write | FTP | Misconfig | Gains user access |
| Guest | Telnet, rlogin | Misconfig | Gains users access |
| Imap | Imap | Bug | Gains root access |
| Named | DNS | Bug | Gains root access |
| Phf | Http | Bug | Executes commands as http user |
| Sendmail | SMTP | Bug | Executes commands as root |
| Xlock | SMTP | Misconfig | Spoof user to obtain password |
| Xsnoop | SMTP | Misconfig | Monitor key stokes remotely |

### 5.4 Network Trace Datasets

Network traces [46] captured from live networks is often used for testing intrusion detection systems. The main advantage of using network traces is that the results do not demonstrate any bias due to artifacts from simulated data that are not representative of actual data. However, when network traces are available, they are often limited. For example, the traces might only contain packet header data, or might be summarized even further into flow-level information.

### 5.5 NetFlow Datasets

NetFlow is used to perform traffic analysis. NetFlow data is aggregated flow level data captured at the router level in online mode. The use of NetFlow data for high-speed traffic analysis is quite important and needed to detect attacks or anomalies from high speed network [47].

However, the use of NetFlow comes at the cost that the packets have been stripped of their content. As Sir Francis Bacon once said: "Knowledge is power" [48]. With this loss of the packet content, some attacks will be impossible to find, because in those cases only the packet content contains relevant information. Still, there is ample scope for detecting attacks or anomalies in high speed network.

Flows are identified uniquely by characteristics of the traffic that they represent, including the source and destination IP (Internet Protocol) address, IP type, source and destination TCP or UDP (user datagram protocol) ports, type of service and a few other items [47]. NetFlow records are sent to the logging host in the following cases.

- For flow representing TCP traffic, when the connection is completed (after a RST or FIN is seen).
- When no traffic for the flow has been seen in 15 seconds.
- 30 minutes after the start of the flow. This causes long lasting traffic patterns to show up sooner than they might otherwise in the log.
- When the flow table fills and meets conditions.

---

[3] http://securityresponse.symantec.com/avcenter

Table 7. Normal and attack traffic information for KDD99 dataset

| Dataset | DoS | | Probe | | u2r | | r2l | | Normal |
|---|---|---|---|---|---|---|---|---|---|
| | Total instances | Attacks | Total instances | Attacks | Total instances | Attacks | Total instances | Attacks | |
| 10% KDD | 391458 | apache2 smurf, neptune, back, teardrop, pod, land, mailbomb, syn flood, ping of death, syslogd, process table udpstorm | 4107 | satan, ipsweep, portsweep, nmap, mscan, saint | 52 | buffer overflow, rootkit, loadmodule, perl, ps, xterm, eject, ffbconfig, fdformat | 1126 | warezclient, guess passwd, warezmaster, imap, ftp write, multihop, phf, spy, named, sendmail, xlock, xsnoop | 97277 |
| Corrected KDD | 229853 | | 4107 | | 52 | | 1126 | | 97277 |
| Whole KDD | 229853 | | 4107 | | 52 | | 1126 | | 97277 |

Table 8. Background and attack traffic information for the LBNL datasets

| Date | Durati-on (mins) | LBNL Hosts | Remote Hosts | Background Traffic rate (packet/sec) | Attack Traffic rate (packet/sec) |
|---|---|---|---|---|---|
| 10/04/04 | 10 min | 4,767 | 4,342 | 8.47 | 0.41 |
| 12/15/04 | 60 min | 5,761 | 10,478 | 3.5 | 0.061 |
| 12/16/04 | 60 min | 5,210 | 7,138 | 243.83 | 72 |

Table 9. Background traffic information for four Endpoints with high and low rates

| Endpoint ID | Endpoint Type | Duration (months) | Total Sessions | Mean Session Rate (/sec) |
|---|---|---|---|---|
| 3 | Home | 3 | 3,73,009 | 1.92 |
| 4 | Home | 2 | 4,44,345 | 5.28 |
| 6 | University | 9 | 60,979 | 0.19 |
| 10 | University | 13 | 1,52,048 | 0.21 |

Table 10. Endpoint attack traffic for two high and two low rate worms

| Malware | Release Date | Avg. Scan rate (/sec) | Port (s) Used |
|---|---|---|---|
| Dloader-NY | Jul 2005 | 46.84 sps | TCP 1,35,139 |
| Forbot-FU | Sept 2005 | 32.53 sps | TCP 445 |
| Rbot-AQJ | Oct 2005 | 0.68 sps | TCP 1,39,769 |
| MyDoom-A | Jan 2006 | 0.14 sps | TCP 3127-3198 |

### 5.6 Discussion

Based on our survey of existing incremental anomaly detection techniques in the context of preprocessed or real-life datasets, we observe the following-

- Of the five datasets reported above, most researchers have established their work using either the Lincoln Lab dataset in offline mode or using the network trace or netflow dataset generated based on their own testbed in online mode.
- Most datasets are useful for offline anomaly detection.
- Some existing datasets are not labeled (e.g., LBNL, Network trace) and do not have any attack statistics (e.g., LBNL, Network trace).
- There is both packet and flow level information in the network trace dataset.

We present a comparison (refer to Table 11) of the existing datasets in terms of different parameters, such as that attack information, and traffic information in terms of normal or attack.

## 6. Evaluation Criteria and Analysis

Benchmark intrusion datasets play an important role in evaluating an attack detection system. But there is only one well known and commonly available benchmark dataset (i.e., KDDCup99) for performance analysis of IDSs. Researchers analyze their detection methods based on live network traffic (i.e., network trace); but they cannot claim that the detection methods work in all situations. Some evaluation criteria are discussed below in brief.

### 6.1 Metrics

The intrusion detection community commonly uses four metrics. The first two are detection rate and false positive rate and, conversely, true and false negative rates [49]. Two other metrics, effectiveness and efficiency are defined by Staniford et al. [50]. Effectiveness is defined as the ratio of detected scans (i.e., true positives) to all scans (true positives plus false negatives). Similarly, efficiency is defined as the ratio of the number of identified scans (i.e., true positives) to all cases flagged as scan (true positives plus false positives) [50], and is the same as the detection rate defined previously.

Table 11. Comparison among intrusion datasets

| Name of the dataset | Categories of attacks | Benchmark (B)/Non-benchmark (N) | Traffic type (normal[N]/attack[A]) | Real/ Non-real time |
|---|---|---|---|---|
| KDDCup99 | DoS, Probe, u2r, r2l | B | Both | N |
| LBNL | - | B | Both | N |
| End-point | Zotob.G, Forbot-FU, Sdbot-AFR, Dloader-NY, So-Big.E@mm, MyDoom.A @mm, Blaster, | N | Both | N |

|  | Rbot-AQJ, RBOT.CCC |  |  |  |
|---|---|---|---|---|
| Network Traces | - | N | N/A | R |
| NetFlow | - | N | N/A | R |

### 6.2 ROC Analysis

Receiver Operating Characteristics (ROC) curve is often used to evaluate the performance of a particular detector. This approach is used by Lincoln Lab for evaluation of anomaly-based detection systems and discussed in detail by McHugh [51]. An ROC curve has false positive rate on its x-axis and true positive rate on its y-axis, thus moving from (0, 0) at the origin to (1, 1). The detection system must return a likelihood score between 0 and 1, when it detects an intrusion in order for the ROC curve to provide meaningful results. The ROC curve can be used to determine how well the overall system performs, the most appropriate threshold values given acceptable true and false positive rates, and to compare different detection systems.

### 6.3 Complexity and Delay Comparison

The training and classification time taken by anomaly detectors as well as their training and run-time memory requirements can be computed using the hprof tool [45]. Contrary to common intuition, complexity does not translate directly into accuracy of an anomaly detector. A delay value of ∞ is listed if an attack is not detected altogether. The detection delay is reasonable (less than 1 second) for all the anomaly detectors we surveyed.

Based on our survey, we observe that most researchers use the first evaluation metric i.e., detection rate, false positive rate, effectiveness, and efficiency to establish their works.

### 7. Research Issues and Challenges

Based on our survey of published papers on incremental anomaly detectors, we observe that most techniques have been validated using the KDD99 intrusion datasets in an offline mode. However, the effectiveness of an ANIDS based on incremental approach can only be judged in a real-time environment. Following are some of the research issues we have identified in this area.

- Most existing IDSs have been found inadequate with new networking paradigms currently used for both wired and wireless communication[4,5]. Thus, adaptation to new network paradigms needs to be explored.
- Most existing methods are dependent on multiple input parameters. Improper estimation of these parameters leads to high false alarm rates.
- The clustering techniques that are used by anomaly detectors need to be faster and scalable when used on high dimensional and voluminous mixed type data.
- Lack of labeled datasets for training or validation is a crucial issue that needs to be addressed. The KDD99 datasets are out-of-date. New valid datasets need to be created and made available to researchers and practitioners. Developing a reasonably exhaustive dataset for training or validation for use in supervised or semi-supervised anomaly detection approaches is a challenging task.
- Estimation of unbiased anomaly scores for periodic, random or bursty attack scenarios is another challenging issue.
- Lack of standard labeling strategies is a major bottleneck in the accurate recognition of normal as well as attack clusters.
- Development of pre or post-processing mechanisms for false alarm minimization is necessary.
- Handling of changing traffic pattern remains a formidable problem to address.

### 8. Conclusion

In this paper, we have examined the state of modern incremental anomaly detection approaches in the context of computer network traffic. The discussion follows two well-known criteria for categorizing of anomaly detection: detection strategy and data source. Most anomaly detection approaches have been evaluated using Lincoln Lab (i.e. KDDCup99 intrusion dataset), Network Traces and LBNL datasets. Experiments have demonstrated that for different types of attacks, some anomaly detection approaches are more successful than others. Therefore, ample scope exists for working toward solutions that maintain high detection rates while lowering false alarm rates. Incremental learning approaches that combine data mining, neural network and threshold based analysis for the anomaly detection have shown great promise in this area.

### Acknowledgement

This work is supported by Department of Information Technology (DIT) and Council of Scientific & Industrial Research (CSIR), Government of India. This work is also partially supported by NSF grants CNS-0851783 and CNS-0958576. The authors are thankful to the funding agencies.

---

[4]Bro (online) http://www.bro-ids.org/
[5]SNORT (online) http://www.snort.org/